# Unclonable anti-counterfeiting labels based on microlens arrays and luminescent microparticles

*Vinay Kumar,[a] Stephan Dottermusch,[a] Ngei Katumo,[a] Bryce S. Richards,[a, b] Ian A. Howard *[a, b]*

[a] *Institute of Microstructure Technology, Karlsruhe Institute of Technology, Hermann-von-Helmholtz-Platz 1, 76344, Eggenstein-Leopoldshafen, Germany.*
[b] *Light Technology Institute, Karlsruhe Institute of Technology, Engesserstrasse 13, 76131 Karlsruhe, Germany.*

Corresponding author: Dr. Ian Howard, email: ian.howard@kit.edu

**Abstract:**

Micron-scale randomness during manufacturing can ensure anti-counterfeiting labels are unclonable. However, this security typically comes at the expense of complex hardware being needed for authentication (e.g., microscopy systems). We demonstrate unclonable labels that can be authenticated using a standard light-emitting diode and smartphone camera. The labels consist of a microlens array laminated to a polymer film that is doped with luminescent microparticles. The micron-scale random overlap of focal volumes and microparticles leads to a pattern of bright points of visible light emission that can be easily imaged by a smartphone camera. 10 000 comparisons of images demonstrate that the labels can be robustly authenticated, and that the probability of a false authentication is on the order of $10^{-15}$. The ability for microlens arrays to simplify the hardware needed for authentication of unclonable labels is generalizable, and attractive for the implementation of unclonable labels in anti-counterfeiting systems.

**Introduction**

Counterfeiting is a growing challenge for industries and governments worldwide. In 2016, the loss to the global economy due to counterfeit products was USD 509 billion (3.3% of the international trade), and it is estimated to reach up to USD 1.9 trillion by 2022 (*2, 3*). Within the European Union (EU), counterfeit products valued at USD 134 billion (6.8% of the EU-import) were traded in 2016 (*2*). Beyond economic consequences, counterfeiting can also endanger human lives, for example, by producing unsafe or ineffective pharmaceutical products (*4, 5*), and compromising safety standards (*6*). To combat counterfeiting, an approach gaining significant interest is to uniquely identify products with unclonable labels, often referred to as physically-unclonable functions (PUFs). Via such an approach, an unclonable label is attached to a product at the point of manufacture and its unique physical properties are characterized, digitized, and stored in an electronic database. The label may also be assigned an easily read serial number, and the digitized physical properties stored alongside this serial number to facilitate lookup in the database. Along the supply chain, actors with the appropriate hardware to characterize the physical properties of an unclonable label can query its authenticity by accessing the database (*1, 7*) – a schematic of this process is shown in the Supplementary Materials, **Fig. S1**. A key advantage of the unclonable label design that we present here, is the simplicity of the hardware needed to characterize the physical properties of the unclonable label. This could reduce the cost of the anti-counterfeiting system and increase its effectiveness by increasing the number of actors (including end consumers) that are able to check the authenticity of a label.

A wide variety of approaches to creating unclonable labels exist. The operational principle, strengths, and weaknesses of a salient selection of examples are briefly introduced in the following and are summarized in **Table S1**. The seminal work of Pappu *et al.* in 2002 demonstrated that the macroscopic speckle-pattern caused by the scattering of coherent radiation from micron-sized beads embedded in a polymer matrix is unique and unclonable (*8*). Shortly after that, Cowburn and coworkers showed that the random surface texture of objects (such as paper documents) can itself enable forgery-proof authentication (*9-13*). These authors developed an apparatus consisting of an excitation laser and several photodiodes whose signals are monitored as a surface (such as a paper document or plastic card) is swept underneath (*9, 10*). Later, work by Toreini and coworkers demonstrated images captured by a standard camera of the transmission of light through a piece of paper are sufficient to uniquely identify the paper (*13*). However, widespread commercial application of these technologies has not materialized due to several drawbacks. The coherent scattering of coherent radiation from micron-sized beads used by Pappu *et al.* is sensitive to the properties of the incident coherent radiation, with the careful control of the spatial coherence, divergence, and incidence angles of the coherent radiation adding technical complexity to the equipment needed for authentication. The maintenance of these characteristics over time on many authentication systems presents a significant practical challenge. The latter two techniques reliant on surface texture and/or paper as a label medium are susceptible to label damage and, in both cases, rather specific hardware configurations and conditions are needed for authentication.

To overcome these challenges, there has been a significant effort in developing novel materials, or material combinations to enable robust unclonable labels. For example, in 2014, Kim and coworkers showed that unique unclonable labels can be made by the random formation of networks of silver nanowires in a polymer matrix, and these could be authenticated under an optical microscope (14). In 2015, Bae et al. demonstrated that unique, unclonable patterns of surface wrinkles are created upon drying of rigid-shell soft-core microparticles and that these can be authenticated using confocal laser scanning microscopy (15). A year later, Zheng et al. used fluorescein-embedded Ag@SiO$_2$ plasmonic nanoparticles to create unique and random emission point patterns (emission from individual particles) that were observable using a fluorescence microscope (*16*). Liu *et al.* demonstrated in 2019 that by introducing PMMA nanoparticles to the ink-receiving layer before ink-jet printing luminescent quantum dots, pinning points are created around which unique micron-scale patterns of the luminescent quantum dots are created during drying (*17*). Carro-Temboury *et al.* recently demonstrated that zeolite cubes on a scale of 10 µm can be loaded with a mixture of emissive lanthanide elements europium, terbium, and dysprosium trivalent ions that exhibit red, green, and blue emission, respectively (*7*). The different colored emissions can be imaged using an optical microscope to reveal the random loading of each zeolite crystal, again leading to an unclonable label (*7*). The same group followed up this work with an excellent example of unclonable unique tags-based imaging random distributions of micron-sized scattering or luminescent particles on a surface, either using a microscope or with a smartphone camera equipped with a clip-on macrolens (*18*). Also recently, Leem *et al.* developed luminescent patterns of fluorescent silk microparticles embedded in a silk film to create edible unclonable fluorescent labels that can be directly attached to pharmaceuticals; these were read out using a scientific camera with a zoom lens and liquid-crystal-based tunable filter (*19*).

Whereas the above-mentioned results greatly expand the application scope of unclonable labels, approaches that allow for simpler hardware to be used for authentication while maintaining or improving the label security are still desirable. As introduced in **Fig. 1**, we demonstrate unclonable labels based on a microlens array (MLA) laminated on a layer doped with luminescent microparticles that meet these challenging requirements. The new labels incorporate micron-scale randomness that guarantees security but is easily observable on the macro scale. Such a label design can be authenticated by a standard smartphone camera, as demonstrated in **Fig 1a**. The working principle is illustrated in **Fig. 1b**, which demonstrates how a brighter emission from a phosphor microparticle occurs when it is the focal volume of a microlens. In the following, we first present a fundamental proof-of-principle study using upconversion (UC) phosphors and laser excitation to demonstrate and characterize the working principle of the MLA-based labels. We then establish, that, as shown in Fig. 1, this label concept can be generalized to normal down-shifting (DS) phosphors and low-cost excitation and detection sources. Finally, we discuss possible directions for further development of this label concept that would further simplify its practical implementation.

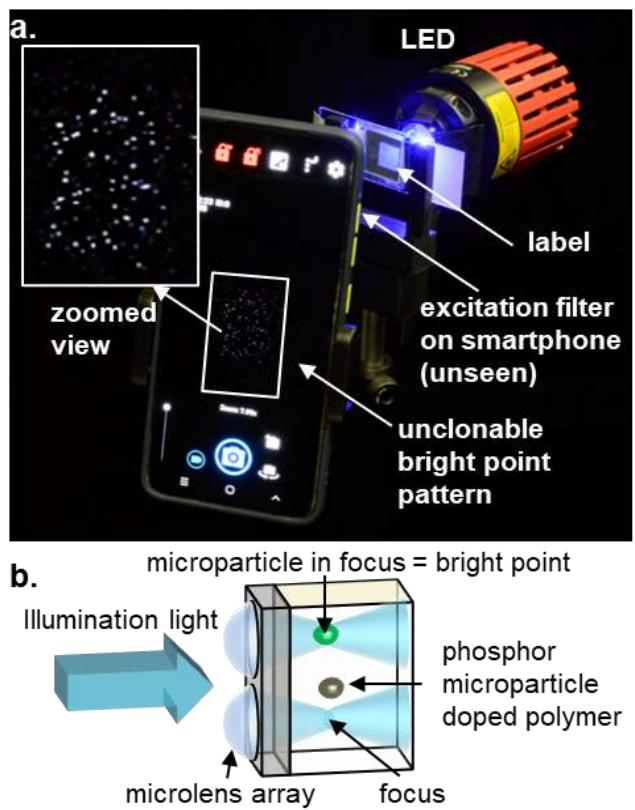

**Fig. 1. Unclonable label concept based on microlens array laminated to polymer layer doped with microparticle phosphors.** a) The unclonable bright point pattern created by micron-scale random overlap of microlens focal volumes with phosphor microparticles can be excited with a blue LED and detected with a smartphone camera. In the current instance, an optical filter is placed before the smartphone camera to reject the excitation light. b) The label consists of a microlens array attached to a polymer layer with phosphor microparticles randomly distributed through its bulk. When a microlens focuses incident light onto a microparticle, much brighter emission is observed from this location. This pattern of bright emission points, related to micron-scale randomness, is easily observable at the macroscopic scale.

**Results and Discussion**

**Labels based on Upconversion phosphors**

In the first instance, we use laser excitation and upconversion phosphors to rigorously test the limits of this label system. Later, we show that LED illumination and standard phosphors are also sufficient for the implementation of these concepts. An MLA is laminated to a polymer layer containing microparticle phosphors. The initial concept for our label and authentication is illustrated in **Fig. 2**. 980 nm laser light impinging on the label through the microlenses (ML) is focused into focal volumes under each ML in the microparticle-doped layer. If a UC microparticle occupies the focal volume under a given ML, bright visible emission will be observed from that point (UC phosphors having the ability to convert invisible NIR radiation to visible radiation in a non-linear manner). Thus, based on the random overlap of the micron-scale particles and focal volumes, a unique pattern of bright emission points is created. **Fig. 2a** shows a schematic illustration of a prototypical implementation of our label design and an example of an authentication system. In these initial tests, a collimated excitation beam is incident on the MLA coated side of the label. The label is held vertically on a rotation stage that controls the angle of incidence (AOI) of the excitation light. A camera is placed behind the label, which observes the back face of the label and captures the emitted light through an optical filter that rejects the excitation light. **Fig. 2b** schematically illustrates the microscopic principle of operation. If a luminescent microparticle happens to lie in the focal volume of an ML, then the emission from this particle is much brighter than from other particles. The emission from the microparticles is isotropic, but a certain fraction, schematically illustrated by the green cones in **Fig. 2b**, will escape the substrate and be collected by the camera.

As presented in **Fig. 2b**, at normal incidence, the ML-foci lie directly on the ML center axis. This leads to a given subset of the ML foci to coincide with microparticles, and the bright point image labeled as pattern 1. For non-normal incidence, the foci move as indicated in **Fig. 2c**. For collimated input light, the lateral shift ($x$) of the foci due to a change in the AOI ($\theta$) is given by $x = F \tan(\theta)$, where $F$ is the focal length of the ML. The change in the position of the foci with AOI is given by $\frac{dx}{d\theta} = F sec^2(\theta)$. For AOIs close to normal $sec^2(\theta) \approx 1$, and $dx \approx F d\theta$. For this initial demonstration, $F$ was 1900 µm so the change in AOI of 2° led to a shift of the foci of 66 µm. This is significantly larger than the 10 µm typical size of the UC microparticles (**Fig. S1**), and therefore a different subset of MLs under which a microparticle is now in the focal volume. This then leads to the new point pattern shown in **Fig. 2c**.

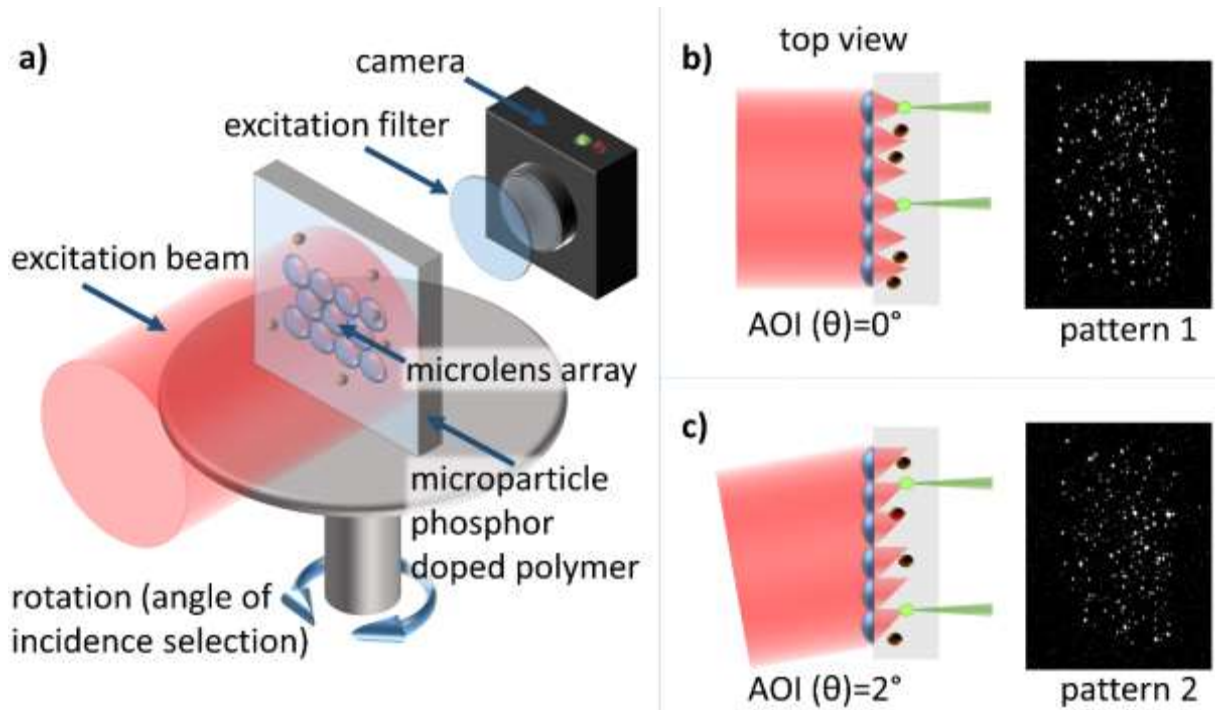

**Fig. 2. Illustration of label concept and example authentication system.** a) A collimated excitation beam impinges on the MLA and is focused into many small volumes in the microparticle-phosphor-doped polymer layer. A camera collects the emission (through a filter to reject the excitation light), which in this first instance is located behind the label. The label is placed on a rotation stage to control the angle-of-incidence (AOI) of the excitation light. Schematic illustrations of the microscopic principle of operation and changing emission point patterns collected at an AOI of 0° (b), and 2° (c) along with example images of the different emission patterns collected from the same label at the respective AOIs.

To summarize the concept, our label design exploits the simplicity with which the micron-scale overlap of focal volumes under MLs and microparticles can be observed (in terms of a bright point pattern) on the macroscopic scale. As the location of the focal volumes depends on the AOI of the light impinging on the MLA, the bright point pattern will change as a function of the AOI. This can be used to enhance the label security by requiring the different bright point patterns at two or more AOIs to authenticate the label. Thus, even the difficult counterfeiting of a single bright point pattern would be insufficient to clone the label, ensuring that these macroscopically read-out labels are unclonable.

**Fabrication of Labels**

We followed the procedure illustrated in **Fig. 3a** to fabricate MLAs in photoresist on glass substrates then laminate them onto polydimethylsiloxane (PDMS) layers doped with micron-sized phosphors. The MLAs were prepared by means of two-photon lithography in a photoresist on a 1 mm-thick glass slide as shown in **Fig. 3b-d** (see Materials and Methods for details). The MLs were designed with a spherical surface (625 µm radius of curvature), leading to a focal length, $F = R\,n/(n-1)$, of 1900 µm. This assures that the focal volume is well contained within the 2 mm-thick PDMS layer. A 250 µm ML base diameter was employed, leading to a ML height of approximately 13 µm. Each MLA consisted of 240 MLs, arranged in a 5 mm x 3 mm hexagonal array with a pitch of 250 µm. Owing to the elastomeric nature of the cured PDMS matrix, optical contact was established between the substrate and the PDMS matrix by pressing the two together (**Fig. 3a**). In the first instance, three of these labels were fabricated for characterization. A 2 mm thick PDMS layer doped with the desired concentration of micron-sized phosphors was prepared (see Materials and Methods for details). The first phosphor chosen was the UC phosphor $Gd_2O_2S$: 18% $Yb^{3+}$, 2% $Er^{3+}$. This was synthesized in-house by flux-assisted solid-state method (*20*), and doped into the PDMS 0.1 wt.%. This UC phosphor that emits in the visible upon excitation at 980 nm was chosen for the initial demonstration as the intensity of its emission can depend on the square of the excitation intensity. This nonlinear dependence increases the contrast between the emission intensity of particles that are in a focus and those that are not. Despite this initial demonstration using favorable UC phosphors, we demonstrate later that the contrast provided by standard DS phosphors is also more than adequate and confirm that more standard phosphor can also be used in this label concept.

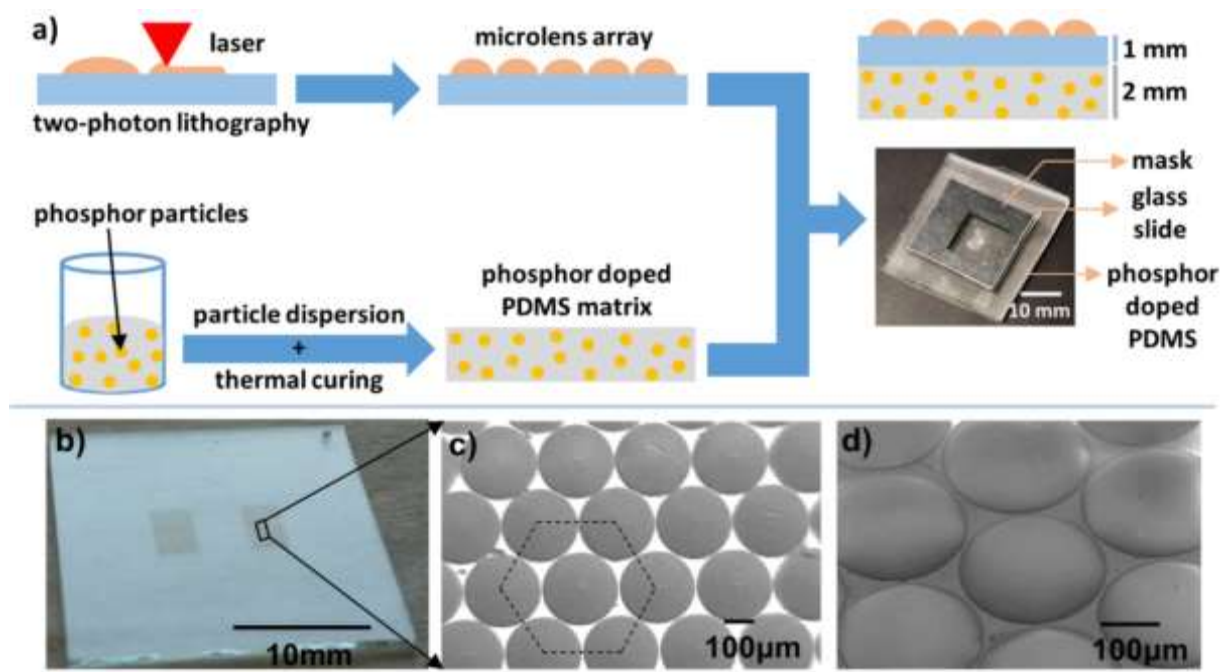

**Fig. 3. Label fabrication** a) schematic of the label fabrication from the lamination of a polymer-based MLA on a glass substrate phosphor-doped PDMS matrix, (b) ITO-coated glass substrate with printed MLAs on the front surface, (c) scanning electron microscope (SEM) image of hexagonally packed MLA on the glass substrate, (d) close up SEM image of a ML

**Label Characterization**

To characterize the prototype UC-phosphor-based labels they were placed in a configuration as illustrated in **Fig. 2a**. The MLs were illuminated using a 980 nm CW laser (SolsTiS, M2). A pair of lenses were used to expand the beam to a diameter of 10 mm. The power density of the excitation beam was 100 mW cm$^{-2}$. The phosphor emission was captured using a scientific CMOS camera (CS2100M-USB, Thorlabs) combined with a zoom lens (MVL7000, Thorlabs). A 900 nm short-pass filter (FES0900, Thorlabs) was inserted before the camera to eliminate the excitation light in the capturing image. **Fig. 4** presents a qualitative analysis of the similarity of the images created for the three following cases:

1) <u>Trials of the same label at the same angle</u>. To demonstrate the reproducibility of the patterns generated by the same label under identical excitation conditions, a single label (L1) was imaged three times under normal incidence illumination. Between each image, the label was removed, the stage rotated to a random angle, then the stage was set back to normal incidence, and finally, the label was replaced on the holder. A grayscale image is acquired from the monochrome camera for each of the three trials. **Fig. 4a** presents these three images, along with a final comparison image in which the three images are placed into the red (R), green (G), and blue (B) channels, respectively, to form an RGB image. This RGB image highlights the reproducibility of the pattern generation. Apart from a few slight fluctuations, nearly all bright points from the three separate images all lie on top of each other, leading to white points in the RGB image rather than separate points of an individual color.

2) <u>Trials of the same label at different angles</u>. To show the influence of the AOI, a single label (L1) was placed into the holder, and images were taken at AOI = 0°, 10°, and 20°. As discussed above, the label should generate different bright point patterns for each AOI. The bright point-pattern images are shown in **Fig. 4b**, alongside a final RGB composite in which each image is put into a single-color channel. In contrast to **Fig. 4a**, no white spots appear, but rather many spots of pure colors. This is a clear representation of the uniqueness of the bright point pattern created at the different AOIs from the same label.

3) <u>Trials of different labels at the same angle</u>. To highlight the uniqueness of different labels (L1, L2, L3) they were placed into the holder one after the other. Every time an image was taken under normal incidence illumination. These grayscale images, as well as the RGB color composite created by placing each of these images into one color channel, are presented in **Fig. 4c**. Just like in **Fig. 4b**, the bright points of single colors indicate that there is no correlation between the images, and that each label is unique.

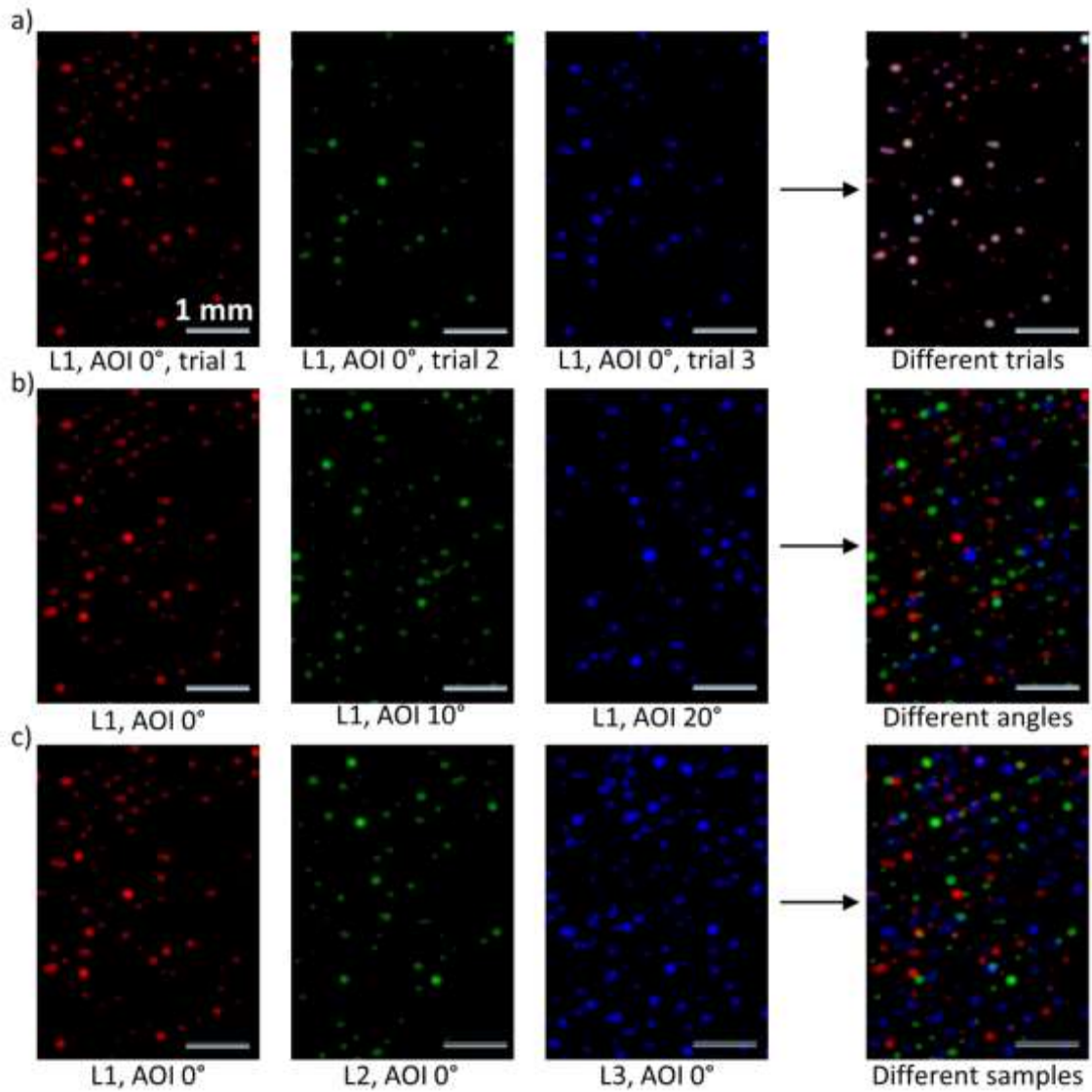

**Fig. 4.** Processed images of the observed point patterns generated by the particles in the ML focal point. Different colored images from: a) the same label (L) taken under the same AOI in different trials of repositioning into the stage; b) the same label under different AOI; and c) different labels taken under the same AOI. The right most images show compositions of the three images to their respective left.

**Label Authentication**

The above analysis qualitatively introduces the features of our anti-counterfeiting label design. However, to understand the capabilities of these labels more quantitatively, authentication by an algorithm is necessary. Several approaches to algorithmic authentication of point patterns exist (*21-23*), which we adapt as follows to our specific situation; namely, to compare the similarity of the point pattern between two images (a "reference image" and a "test image") in a way that is invariant to affine transformations. A detailed description of the algorithm is presented in the Supplementary Materials (section S2). Briefly, the algorithm compares the reference image and test images by:

1) *Processing an image into a list of points.* Each image, be it a reference or a test image, is processed to generate a list of points as follows. First, the images are binarized. This was done by finding the noise level of the image in a region outside the MLA, then assigning a value of '1' to pixels with a value greater than three times this noise level, and zero to all remaining pixels. From the binary image, contiguous bright regions are detected and enumerated. After this, the indices $(x, y)$ of the pixel at the center of each bright spot are determined. Finally, the binarized image is used to mask the original image, then for each bright spot, the intensity of all the pixels within the spot in the original image are summed. In the end, this processing generates a sorted list of points ($\mathcal{P}$), in which the first point ($p_1$) is the indices for the brightest spot and the last point ($p_P$) are those of the dimmest.

2) *Comparison of a test image against a reference image.* In the comparison phase of the algorithm, the list $\mathcal{P}$, obtained from the reference image, is compared to the list $\mathcal{Q}$, obtained from the test image. The following consideration was made: If $\mathcal{P}$ and $\mathcal{Q}$ originated from the same label, illuminated under the same incidence angle, an affine transformation should exist that makes $\mathcal{P} = \mathcal{Q}$. The search for such an affine transformation was adapted from Wolfson *et al.* (*24*). We implement this concept by finding many possible affine transformations for $\mathcal{Q}$, of which one should be the real transformation that cause $\mathcal{Q}$ and $\mathcal{P}$ to overlap if the test image is from the authentic label. For each affine transformation, a vote is cast for each point in $\mathcal{P}$ that has a point in the transformed $\mathcal{Q}$ list within a threshold distance. The affine transformation with the most votes is considered the most suitable transformation, and this maximum number of votes is recorded and used to establish authenticity. If the number of votes is above a threshold, the reference and test images are considered to match. More details on how the affine transformations are found and applied can be found in the Supplementary Materials.

To generate a set of 100 reference images, four different labels were illuminated from 25 AOIs ranging from −12° to 12° in 1° steps (the labels were of an identical design to those described in the previous section; they were based on the same MLA design and UC phosphor-doped PDMS layer). A 1° should shift the foci by around 38 µm, significantly larger than the 10 µm dimension of the UC phosphor. Thus, each of these images should be uncorrelated (this is rigorously confirmed in the following section).

Then, to generate a set of 100 test images, each label was reinserted into the holder and imaged again under the same AOIs. This dataset allows 10,000 comparisons to be made, of which 100 should authenticate (since the test and reference images were of the same label under the same AOI), and 9900 should not.

We applied the above algorithm to determine the number of votes cast when each test image was compared to each reference image. The following specific conditions were used in the algorithm. Firstly, we truncated the points list to the brightest 32 points in each image for the lists $P$ and $Q$. This number is found to allow robust authentication while decreasing computational cost. Secondly, possible affine transformations were determined by taking the five brightest points in $P$ and $Q$ as basis points. This number proves large enough to allow robust authentication while limiting the number of possible affine transformations to 200 (see Supplementary Materials for details). Thirdly, the threshold distance for casting a vote was considered 15 pixels, corresponding to half the spacing of the MLs in the images. **Fig. 5a** presents the number of votes cast for each reference image compared to each test image. On the diagonal from the upper left to the lower right corner when the test and reference image should authenticate from 24 to 30 votes are cast. For off-diagonal comparisons, a much smaller number of votes, 4 to 9, are cast.

**Fig. 5b** shows the distribution of the histograms of the number of votes cast for the 9900 non-authenticating image comparisons and the 100 authenticating image comparisons. There is a distinct separation between the number of votes cast in the matching and the non-matching cases. We now wish to select a threshold number of votes above which test image will be considered authentic. Based on an analysis of these distributions, we can determine the probability of a false positive and a false negative when an arbitrary threshold number of votes is selected. To do this, the distributions of the number of votes for non-matching and matching images were fit to Gaussian distributions (dashed lines in **Fig. 5b**). The solid lines in **Fig. 5b** exhibit a cumulative distribution function (CDF) for non-matching images to obtain <u>more</u> than a given number of votes and the CDF for matching images to obtain <u>less</u> than a given number of votes. The curves cross at just under 14 votes. To minimize the probability of a false positive or false negative authentication, one would, therefore, choose to take a threshold of 14 votes. The probability that two non-matching images will generate more than 14 votes (giving a false positive match), and that the probability that two matching images have fewer votes (giving a false negative) is on the order of $10^{-15}$. Therefore, we concretely demonstrate that reliable authentication based on labels of our design is possible.

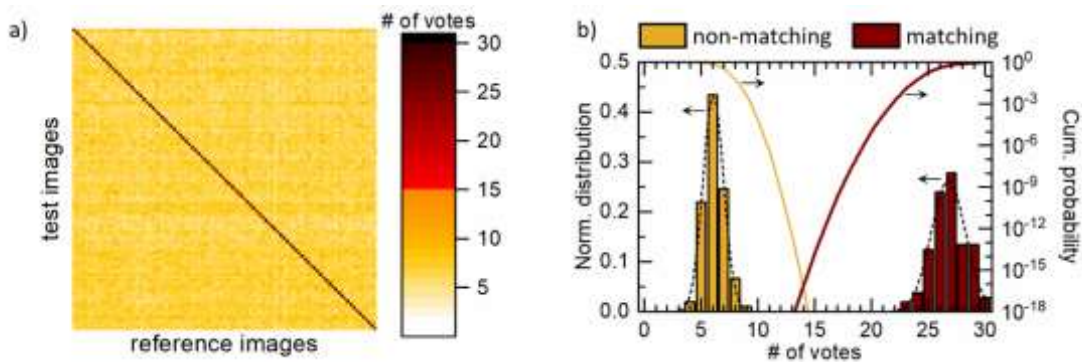

**Fig. 5. Label authentication.** (a) Maximum number of votes obtained with the used algorithm for 100 reference and 100 test images in a one-to-one comparison. (b) Distribution of the number of votes obtained for the 'same' (the same label and same angle) and 'different' (different label or different angle) reference and test images. The solid lines show the cumulative distribution function (CDF) for each distribution, calculated based on the Gaussian fits (dashed lines).

**Authentication as a function of AOI**

To examine in detail the angular tolerance of authentication of the labels discussed in the previous two sections, images were obtained for a single label at AOIs ranging from 0° to 10° in 0.1° steps. The affine-transformation-based algorithm described in the previous section was used to compare images to one another. An image, taken at a certain AOI, was used as the reference image, and all 21 images taken within an angular offset range of ±1° of the AOI image in 0.1° steps were used as test images. The resulting number of votes in these comparisons is plotted in **Fig. 6a**. Using the authentication threshold of 14 votes, we use this data to plot the probability of authentication as a function of the angular offset in **Fig. 6b**. As seen, reproducible authentication is only possible when the absolute angular offset is less than 0.2°. On the positive side, the rapidly changing pattern with AOI leads to high security and many unique patterns. On the negative side, this necessitates excellent control of the AOI between the label and authentication apparatus that is challenging to maintain in the field. The main reason for this rapid change in pattern is the large discrepancy between the ML focal length of ≈1900 µm and the particle dimensions of less than 10 µm. A 0.2° change in the incidence angle corresponds to a 7 µm change in the focus positions. This limitation in the range of AOIs within which label authenticates can be lifted with appropriate engineering of ML characteristics and particle sizes.

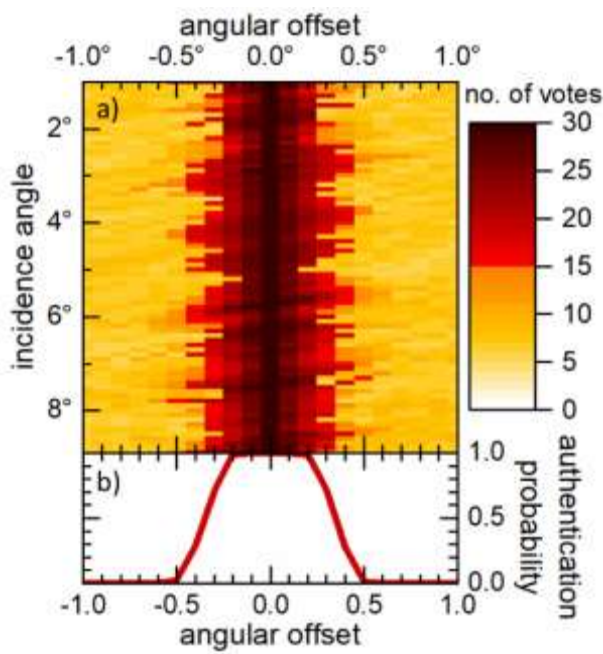

**Fig. 6.** a) Number of votes obtained using the analyzation algorithm when comparing images obtained at various incidence angles to images obtained using an angular offset. b) Probability of authentication as a function of the angular offset between the reference and test images (based on a threshold of 14 votes)

Moving to significantly shorter focal length lenses would be possible by embossing the MLA directly into the micro-particle containing layer. Combined with larger microparticle dimensions, authentication within an AOI range of several degrees should be possible.

**Labels based on standard phosphors**

The results presented in the preceding sections utilized UC phosphors in the label design. These labels required excitation from a high-intensity source, and an expensive scientific camera with a zoom lens was used. In this section, we demonstrate that labels based on standard DS phosphors allow authentication using an inexpensive light-emitting diode source and a standard smartphone camera.

To prepare these labels, the same MLAs and PDMS layers were used as before, but the PDMS was doped with commercially available particles (YYG 557 230 isiphor) at a concentration of 0.5 wt.%. This DS phosphor can be excited with a blue LED (excitation peak at 450 nm) and emit in a broad visible wavelength range of 470 nm to 700 nm. The particles have a spherical shape and a D50-particle size of 30 – 35 µm. The spectral response and SEM images of the phosphor are presented in the Supplementary Materials (sections 3 and 4).

**Fig. 7a and 7b** present the images (after binarization) taken simultaneously with the excitation filtered scientific CMOS camera and the smartphone (Samsung galaxy A71). The contrast between the emission intensity of DS microparticles inside versus outside an ML focus is sufficient that a bright point pattern is clearly resolved. Also, the smartphone and scientific CMOS camera record the same bright point pattern (experimental setup described in Supplementary Material section 5). For a quantitative comparison, passing the images presented in **Fig. 7a**, and **7b** into our authentication algorithm as the reference as test respectively yields 27 votes to be cast in the comparison (far greater than the threshold of 14 votes), indicating that the algorithm also considers these images to be an identical match. In order to visually compare the equivalency of the camera images, the affine transformation related to this largest number of votes was applied to the smartphone image and then a false-color image of the CMOS captured image (green) together with the transformed smartphone image (magenta) was generated and is presented **Fig. 7c**.

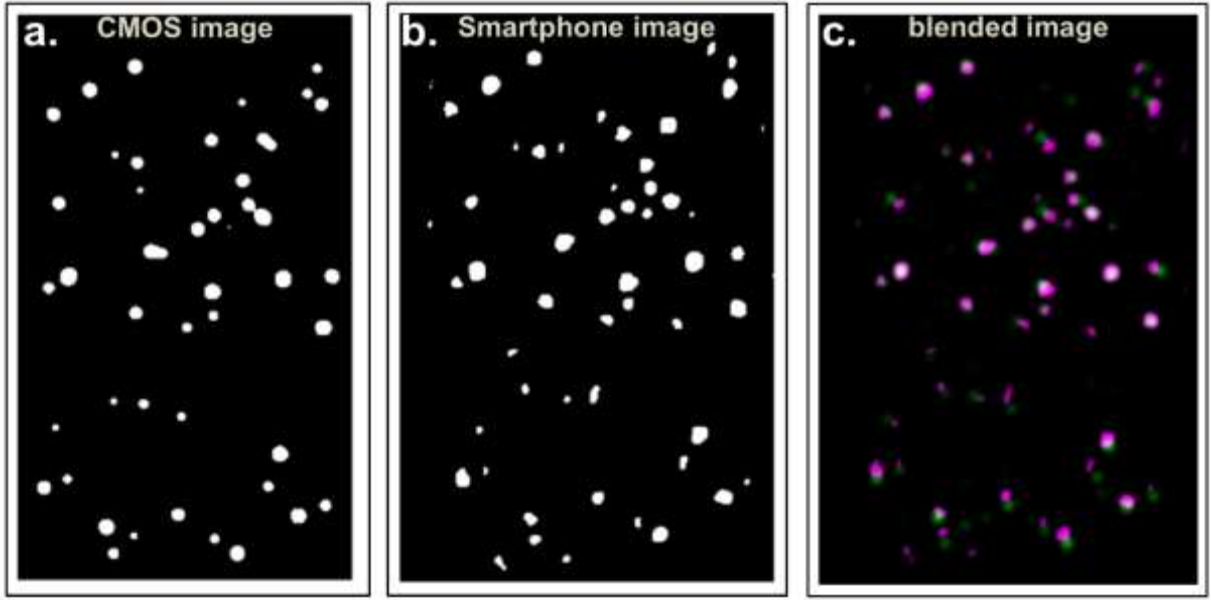

**Fig. 7. Demonstration that an unclonable label based on standard phosphor allows authentication with LED excitation and smartphone detection.** a) DS phosphor pattern observed by the s-CMOS camera; b) DS pattern observed by a smartphone camera c) Composite zoom-in image of CMOS (reference) and smartphone-captured image in a single frame. After registration with an affine transformation.

Thus, we demonstrate that this attractive unclonable label concept can be realized with standard commercial phosphors and use inexpensive light sources and detectors. However, to move further towards the attractive goal of ubiquitous authentication requiring only a single smartphone certain challenges remain. For example, now the labels are measured in transmission geometry. It would be attractive if the flash and camera of a single smartphone could be used to authenticate the label. Redesigning the MLA spacing, and addition of scattering structures around the MLs should allow a single smartphone to provide the illumination and detection source for authentication. Furthermore, an optical filter in front of the smartphone camera is currently needed to block the excitation light. To remove this requirement, we suggest that persistent phosphor microparticles could be used. In this case, they would be excited with a single flash from the LED, then the bright point pattern of the persistent luminescence captured after the excitation light is turned off. Finally, the dependence of the bright point pattern on the AOI means that quite precise control over the label relative to the illumination is currently required. At the level of label design, moving to shorter focal length MLs (directly embossed into the microparticle layer) would help alleviate this, as would some deliberate broadening of the ML foci through lens design. Also, this could be addressed at the anticounterfeiting system level. The labels could be characterized at many AOIs, and then users could simply provide images at several (unknown) AOIs for authentication. If a match is found for each of the images provided by the user, then the label must be legitimate. This would add complexity to the initial label characterization but simplify authentication by the end-user.

**Discussion**

Anti-counterfeiting labels have been realized using MLAs laminated onto microparticle-doped PDMS layers. The overlap of focal volumes with a random subset of microparticles in the layer leads to a pattern of bright emission points. We demonstrate from an analysis of 10 000 image comparisons based on UC phosphor labels that the labels can be robustly authenticated, and the probability of a false positive authentication on the order of $10^{-15}$. The unclonable nature of the labels is based on the random micron-scale overlap of the ML foci and microparticles. The security is further enhanced by the change of the bright point pattern with the AOI of the excitation light. Whereas it might be possible, although challenging, to reproduce a single bright-point pattern by inkjet printing, there is no way that two or more unique bright point patterns at different AOI be reproduced. Thus, requiring two or more patterns to match at different AOI would make these labels highly secure, but still easily authenticated with inexpensive hardware (i.e., a standard blue-emitting LED for excitation, and a smartphone for detection). Refinement of this concept should allow a single smartphone to provide both the excitation, and the detection.

**Materials and Methods**

**Fabrication of microparticle layers:**

Firstly, the desired weight percentage of phosphor particles was dispersed in the silicone elastomer base (SYLGARD 184, Dowsil, RTV-A). A high-speed dispersion device (CAT M. zipper GmbH) was used to disperse phosphor particles uniformly in the RTV-A solution. Then, the curing agent (RTV-B) was mixed throughout the solution, with a component ratio of 10:1. The resulting solution was kept in a vacuum desiccator to extract air bubbles from the mixture.

The mixture was then poured in a 2 mm thick polished brass mold and cured at 150℃ for 30 min in an open-air environment.

The UC phosphor used was gadolinium oxysulfide doped with a ytterbium sensitizer (980 nm absorption) and a visible erbium emitter ($Gd_2O_2S$: 18% $Yb^{3+}$, 2% $Er^{3+}$). It was prepared in-house via flux-assisted solid-state method as described by Katumo *et. al.* (*25*). The characteristic particle size of UC phosphor is in the order of 10 μm and exhibits rod-like morphology as demonstrated in **Fig. S1**. The DS phosphor was the commercial YYG 557 230 isiphor (Merck). This is a cerium-activated aluminum-garnet yellow phosphor, in a spherical shape with a D50 particle size between 30-35μm as demonstrated by the SEM image shown in **Fig. S5** (Supplementary Materials).

**Fabrication of MLA**

The MLA was designed in Matlab (MathWorks) and written using a Photonic Professional GT (Nanoscribe) on a 1 mm thick glass substrate coated with indium tin oxide (ITO). The resist used to fabricate MLA was IP-S (Nanoscribe). Since the refractive indices of IP-S resin and the glass substrate are nearly the same, it is difficult for the device to find the interface print position between the glass and IPS. Therefore, it is recommended to use ITO coated glass substrate with the IP-S photoresin for the 3D micro-texture writing. We printed 240-elements MLA which are distributed in a hexagonally close-packed manner. The MLA took roughly 30 hours to print.

**Hardware setup and experiment**

For observing the visible UC emission, a full-high-definition (HD) CMOS camera (CS2100M, Thorlabs) with an 18-108 mm zoom lens was used (MVL7000 - 18 - 108 mm EFL, Thorlabs) set to a focal length of 108 mm. The camera was placed behind the polymer sample, facing the back surface of the label at a distance of 32 cm. The camera was placed slightly off normal to avoid transmitted laser light and a 900 nm short-pass filter (FES0900, Thorlabs) was used to eliminate light scattered from the laser excitation. To illuminate the MLA surface area with the collimated and expanded version of the laser beam, the beam profile of the excitation beam was expanded using a pair of lenses having focal lengths –25 mm and +100 mm. The label was mounted on a rotational stage, which was fixed on an X-Y stage. The illumination area on the label i.e., the MLA region was made to place at the center of the rotational axis so that each microlens faces the same AOI while rotating the label stage.

In the smartphone readable setup, the camera is covered with a 500 nm long-pass filter (FEL0500, Thorlabs) for eliminating the excitation light in the capturing image. The smartphone is fixed slightly off-axis to the label and the excitation source to avoid the direct illumination from the excitation beam towards the camera. The distance of the smartphone while capturing the emission-based image from the label is about 10 cm. For conducting the smartphone-based authentication experiment, we use Samsung galaxy A71.

**Smartphone camera settings for the image acquisition**

An open-source camera app named Open Camera (v1.48.1 code 77), available in the Google Play store is used to capture the label's bright pattern from an android smartphone. The app provides a user interface platform called Application Programming Interface (API) to manually control the functions of the smartphone camera. For this, the default setting (original camera API) has been changed to a manual setting, called 'Camera2API'. In this manual mode, we have changed the camera settings to capture the label pattern as follows: the spatial resolution is set to 4032*3024 pixels. The focus distance is set to 10cm, the white balance is set to fluorescent, the color effect to none, and then the auto level to unchecked. The focus region of the bright point pattern on the label is digitally zoomed in 8x times to the smartphone screen. The ISO is manually set to 800 and locked at this value to lower the noise in the capturing image. The exposure time is set to 1/30 sec and locked. All these settings are saved and can be retrieved during the time of re-use in the setting manager of the app.

**Acknowledgments**

The authors gratefully acknowledge the Helmholtz Association for funding through HEMF, KNMF, and the MTET program, and the recruiting initiative of B.S.R. The authors V.K. and N.K. acknowledges the DAAD for financial support. The authors thank Dr. Arndt Last at Institute of Microstructure Technology, KIT for his technical support with the photograph in Fig. 1.

**Author Contributions**

V.K. and S.D. fabricated labels, performed experiments, and wrote the authentication algorithm. N.K. synthesized the UC phosphors. All authors contributed to the analysis of the data and writing of the paper. I.A.H. conceived the project.

**Competing Interests**

The authors have filed a patent application and are pursuing commercial application of this technology.

**Materials and Correspondence**

Material requests and correspondence can be addressed to ian.howard@kit.edu.

**Supplementary Information**

Supplementary material for this article is available:

Schematic of unclonable labels in supply chain

Details on image processing algorithm

Phosphor characterization

Experimental details for smartphone / CMOS camera comparison

**Supplementary Material**

**Unclonable anti-counterfeiting labels based on microlens arrays and luminescent microparticles**

*Vinay Kumar, [a] Stephan Dottermusch, [a] Ngei Katumo, [a] Bryce S. Richards, [a, b] Ian A. Howard ∗, [a, b]*

[a] *Institute of Microstructure Technology, Karlsruhe Institute of Technology, Hermann-von-Helmholtz-Platz 1, 76344, Eggenstein-Leopoldshafen, Germany.*
[b] *Light Technology Institute, Karlsruhe Institute of Technology, Engesserstrasse 13, 76131 Karlsruhe, Germany.*

*Corresponding author: Ian A. Howard. Email: ian.howard@kit.edu

**Table of Contents**

## Section 1: Securing products against counterfeiting in supply-chain systems using physical unclonable labels

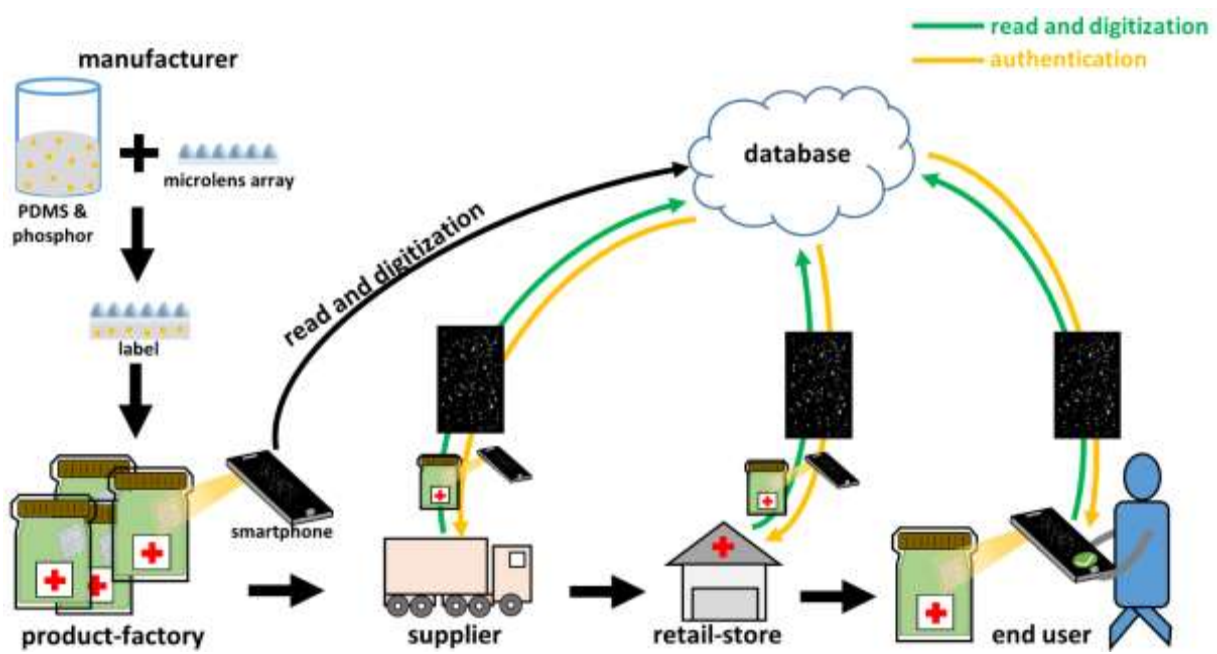

*Fig. S1. Schematic presentation of the label generation and its possible application to product-authentication in a supply-chain system using physical unclonable labels (PULs). The schematic describes an easy and quick authentication of labelled product using a flashlight and camera of a smartphone, at different transfer stages of the supply chain. A similar anti-counterfeiting practise has been illustrated by Gu et al(1).*

Figure S1 shows a schematic illustration of label generation and its possible application to secure the products against counterfeits in a supply chain. The products are secured against counterfeiting by labeling them with physically unclonable function (PUFs) labels, which allows independent authentication of the products at the different stages of the supply chain. In the schematic shown in Fig. S1, a smartphone flashlight and camera are implemented for authenticating the labels. The smartphone flashlight illuminates the unclonable labels and the smartphone camera captures the reflection-based luminescent pattern, therefore has the capability to omit the requirement of a separate excitation source or a separate detector (unlike the requirements for transmission-based label authentication). Although we have read and authenticated our labels via transmission-based PL imaging, it is to be emphasized that our label design technology can also work in reflection mode pattern imaging using a single device (smartphone) for excitation as well as detection.

A digital platform 'database' is used for storing the encoded information for the product. Two schemes are possible, either a test image taken along the supply chain is compared to all reference images in the database, or a unique identification code is provided with every label, thus a test image would only be compared to one reference in the database that matches the code.

**Operational principle, strengths, and weaknesses of Unclonable labels**

*Table S1: List of unclonable label technologies by category including technologies based on unclonable labels, authentication procedure, and limitations with respect to the proposed technology*

| Label | Authentication | Limitations | Reference |
|---|---|---|---|
| Silica microsphere doped epoxy matrix | Macroscopic speckle-pattern caused by scattering of coherent radiation from the microspheres | Extremely sensitive of pointing, collimation, and spatial coherence of the incident radiation as well as the position of the detector | R. Pappu et. al.(*8*) |
| Unique surface of paper documents | Reflected or transmitted intensity profile from the random surface texture of the labelling object | • Careful maintenance of the label over time<br>• Specialized read-out equipment required. | R. P. Cowburn et. al.(*10*) |

| Unique transmission of paper | Transmission image of paper using a back illumination and a standard camera | • Susceptible to damage the label in external environmental conditions<br>• Not reliant on micron scale features | Toreini et. al.(*26*) |
|---|---|---|---|
| Networks of fluorescent dye-coated silver nanowires casted on a polymer matrix | Fluorescence microscope image using a wavelength selected excitation source | An optical microscope as a detector is a mandatory part of the authentication system to investigate the distribution of nanowires in the labelling object. | J. Kim et. al.(*14*) |
| Surface wrinkle micropattern of silica-coated alkoxysilane microparticles co-doped with rhodamine-B | Inspection of pattern of surface wrinkles by confocal laser scanning microscope setup, as well as with a smartphone-clipped portable microscope | To verify the self-assembled micro-pattern by the end-users, additional lens-system is required to build a portable microscope with a smartphone camera. | H. J. Bae et. al.(*27*) |
| Fluorescin-embedded silver@silica plasmonic nanoparticles (NPs) on a lithography defined micropattern | Multilevel authentication by using dark-field, fluorescence, or Raman microscope (to investigate the localized surface plasmon resonance, fluorescence or Raman signals encoded in the core-shell NPs | Various microscope tools required | Y. Zheng et. al.(*28*) |
| Inkjet printable fluorescent quantum dots pinned on a Poly-methylmethacrylate (PMMA) film | Examination of microscopic flower-shaped luminescent patterns created by ink | Microscope required | Y. Liu et. al.(*29*) |

| | drying around the pinning points | | |
|---|---|---|---|
| Lanthanides (III) doped zeolite taggants in Polyvinyl alcohol (PVA) polymer | Can use confocal fluorescence microscope (for excitation-selected imaging), and using a CCD-based spectrometer (for emission based spectral imaging) | A dedicated optical microscopic system is required to authenticate the fluorescent labels. | Carro-Temboury et. al.(*7*) |
| Scattering, absorbing and luminescent microparticles doped polymeric film coated on a QR-code sheet | Authentication of randomly distributed microparticles pattern using a smartphone | Additional optics (macrolens) is required with a smartphone camera, to authenticate the microparticles pattern. | Arppe-Tabbara et. al.(*18*) |
| Protein fused silk luminescent microparticles as an edible PUF label on pharmaceuticals | Authentication of luminescent pattern by using a filter-in CCD camera equipped with a zoom lens, or with a filtered flashlight and camera of a smartphone | Luminescent pattern of silk microparticles is indistinguishable by the naked eye. | Leem et. al.(*30*) |

## Section 2: Image processing algorithm

The full code of the image processing algorithm alongside the images used to generate Figure 4 and figure 5 is available at GitLab repository (project id 25605, https://git.scc.kit.edu/zl3429/algorithm_label_authentication/-/tree/master). The algorithm is summarized in Fig. S2.

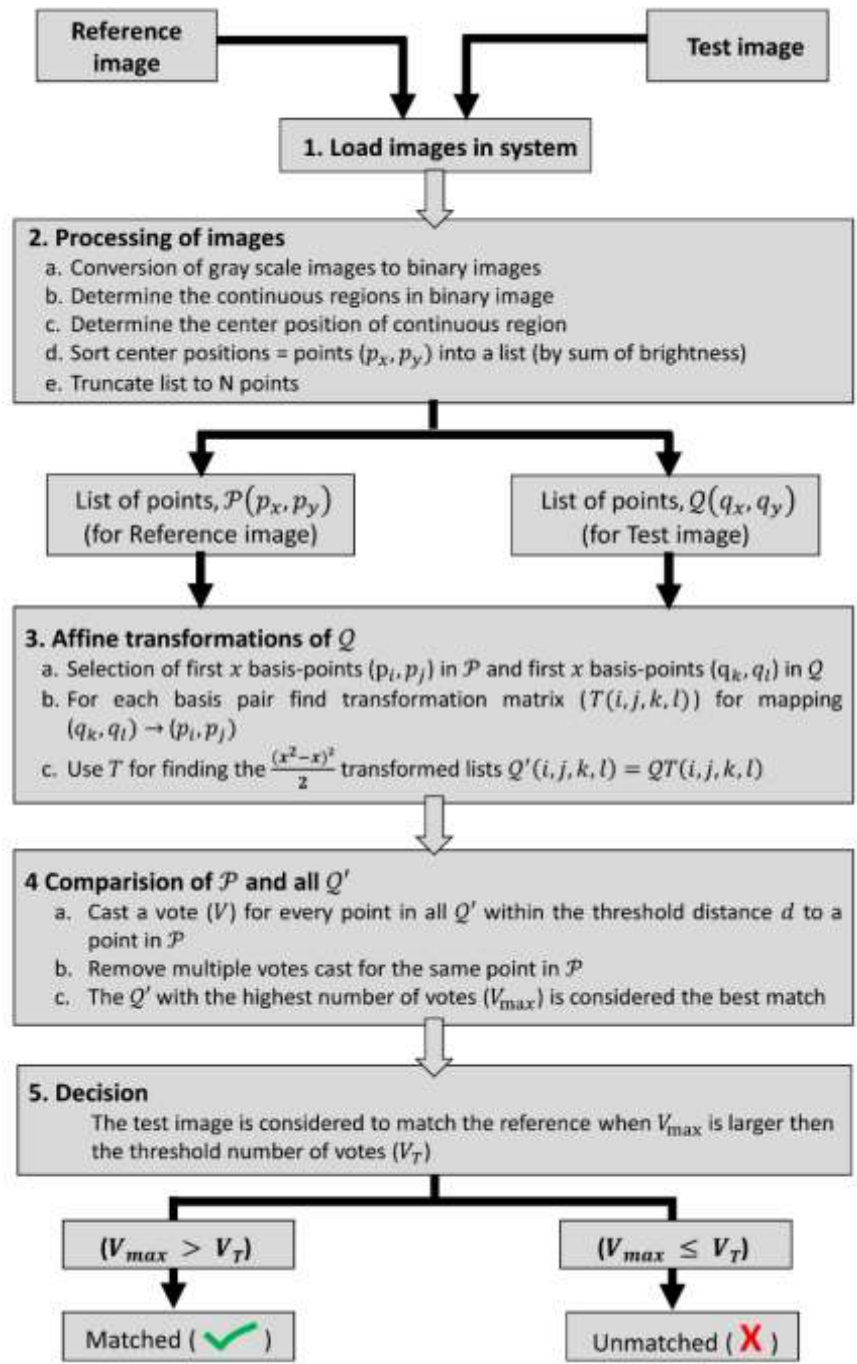

Fig. S2. Flow chart of algorithm for testing and authenticating a label

**Affine transformation**

The affine transformations for the pattern validation are determined as follows: $x$ brightest points in both $Q$ and $P$ are considered. For each of the $x(x-1)$ permutations of the ordered choice of two points from $x$ brightest points in $\mathcal{P}$, an affine transformation can be found that maps the location of those two points, $(p_i, p_j)$, onto the positions of each of the $x(x-1)$ possible permutations of the $x$ brightest points in $\mathcal{Q}$. This leads to $(x^2 - x)^2$ affine transformations that can be applied to $\mathcal{P}$, before it is compared with the reference point list $\mathcal{Q}$. As the transformation $(p_i, p_j) \rightarrow (q_i, q_j)$ is equivalent to $(p_j, p_i) \rightarrow (q_j, q_i)$, then only half of the affine transformations are unique and need to be considered. Thus $\frac{(x^2-x)^2}{2}$ comparisons are necessary to determine how well the two lists match, in a way that is invariant to affine transformations between the reference and test images on which the point lists are based. In our case, $x = 5$ proved reliable for finding matching basis pairs, while at the same time limiting the number of necessary comparisons (to 200).

**Section 3: Experimental section**

**Up-conversion (UC) material: Diffuse reflectance and photoluminescence (PL) spectrum**

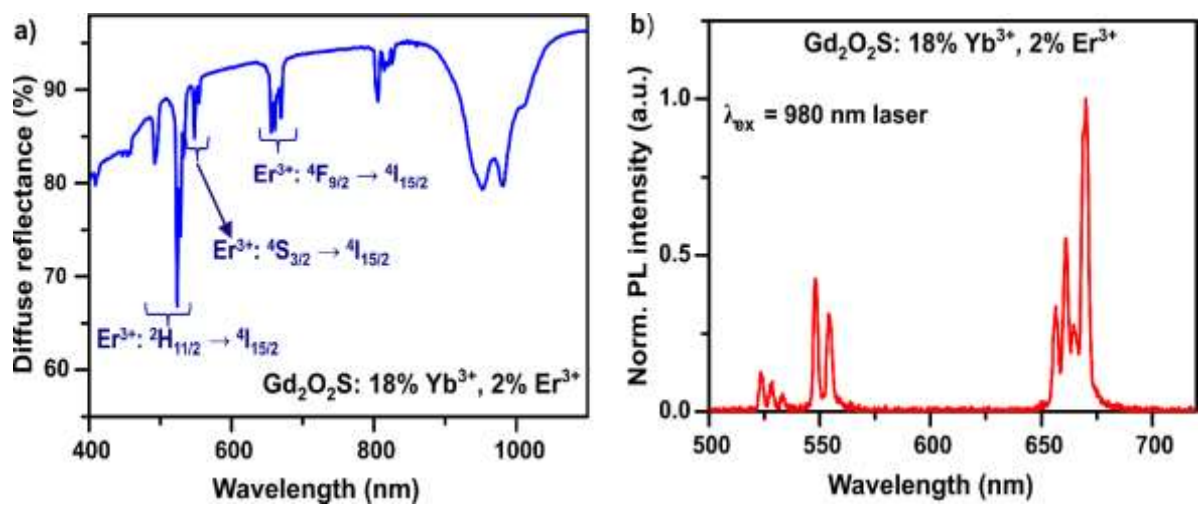

*Fig. S3. a) diffuse reflectance; and 2b) emission spectra of Gd$_2$O$_2$S:Yb$^{3+}$,Er$^{3+}$ up-conversion (UC) phosphor*

Figure S3 shows the diffuse reflectance and up-conversion (UC) emission spectra of gadolinium oxysulfide (Gd$_2$O$_2$S:18%Yb$^{3+}$,2%Er$^{3+}$) phosphor. The diffuse reflectance spectra is recorded over the wavelength range 400-1100 nm using an ultraviolet (UV)-visible-near-infrared (NIR) spectrophotometer (Perkin Elmer Lambda 950) equipped with an integrating sphere. The emission spectra is measured using a compact spectrometer (CCS100/M, Thorlabs) under the excitation of 980 nm laser beam.

**Downshifting (DS) phosphor: excitation and photoluminescence spectra**

Figure S4 shows the excitation and emission spectra of the commercialized downshifting (DS) phosphor (YYG-557-230 isiphore, Merck). The measurement is performed using a spectrofluorometer (Varian Cary Eclipse).

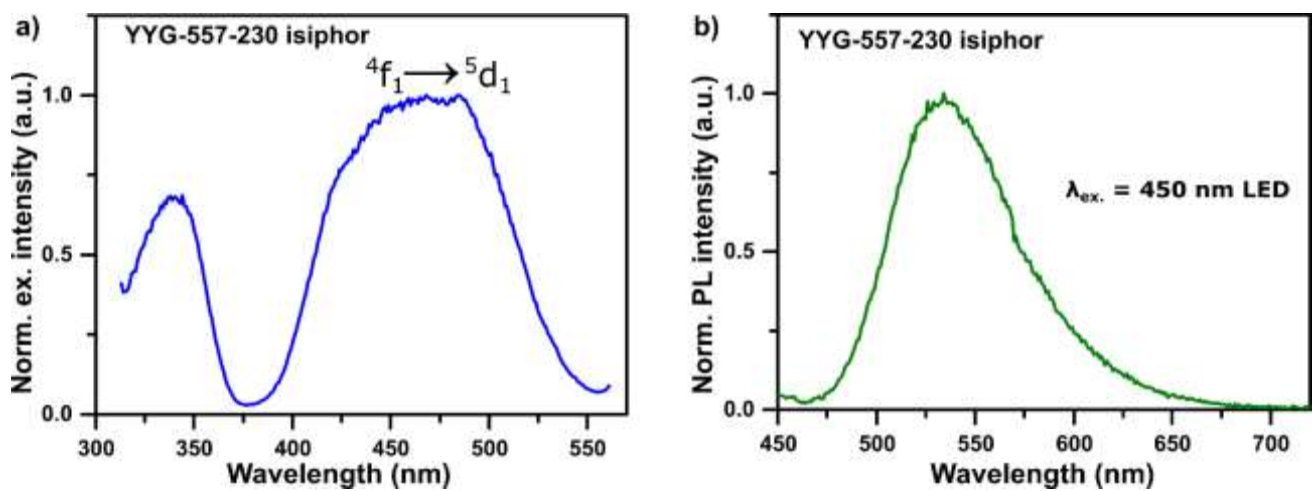

*Fig. S4. Excitation and emission spectra of the DS phosphor (YYG-557-230 isiphore)*

.

**Downshifting phosphor: Diffuse reflectance spectra**

The diffuse reflectance spectra of commercialized DS phosphor (YYG-557-230) is measured over the wavelength range 300-900 nm using a UV-VIS-NIR spectrophotometer (Perkin Elmer Lambda 950) equipped with an integrating sphere.

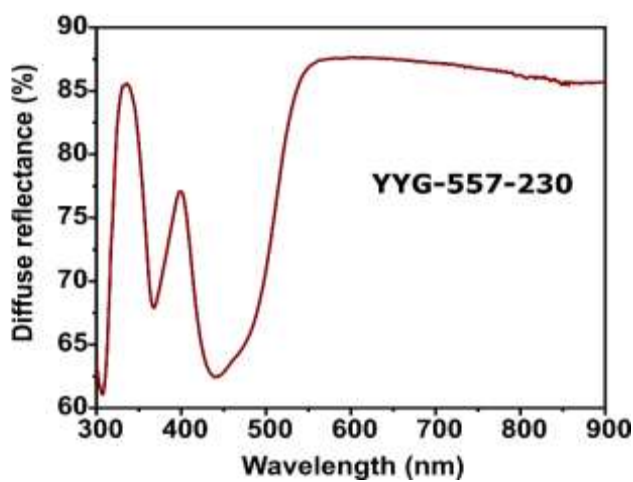

*Fig. S5. Diffuse reflectance spectra of standard DS phosphor (YYG-557-230)*

## Section 4: Scanning electron microscope

The size and structural morphologies of the UC and DS phosphors are studied using the Scanning electron microscope (SEM, Zeiss Supra 90). The SEM confirms that the $Gd_2O_2S$ UC particles are in rod shape, whereas the commercialized DS particles are in a spherical shape. The volumetric size of UC particles is ~ 10 µm, and the DS particles have an average D50 in the 30.5-34.5 µm range.

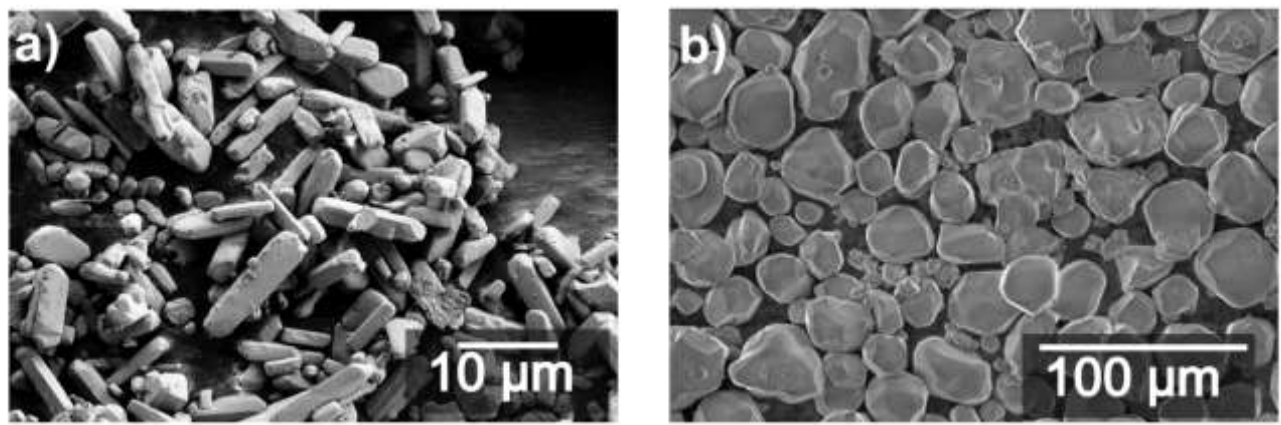

*Fig. S6. a) SEM of $Gd_2O_2S$ UC particles; 5b) SEM image of the commercialized DS phosphor (YYG-557-230)*

## Section 5: Pattern authentication setup using a smartphone and S-CMOS camera

Fig S7 shows the schematic setup used to capture the DS emission pattern from the label. The DS label was illuminated with a 450 nm mounted LED (M450LP1, Thorlabs) and the emission-based images were captured using both a CMOS camera (as a reference image) and a smartphone camera (Samsung galaxy A71) facing the back surface of the label. Although the LED emits strongly divergent light, as the LED is placed 10 cm from the label the light reaching an ML of hundreds of micrometers dimension can be considered collimated. For conducting the experiment with the DS phosphor doped label, the excitation intensity from 450 nm LED was choose at 10 mWcm$^{-2}$. A 500 nm long-pass filter (FEL0500, Thorlabs) was inserted before the smartphone and scientific camera. The smartphone was placed 10 cm away from the label to capture the emission pattern of the label. Whereas the S-CMOS camera is equipped with a zoom lens and is placed 32 cm behind the label.

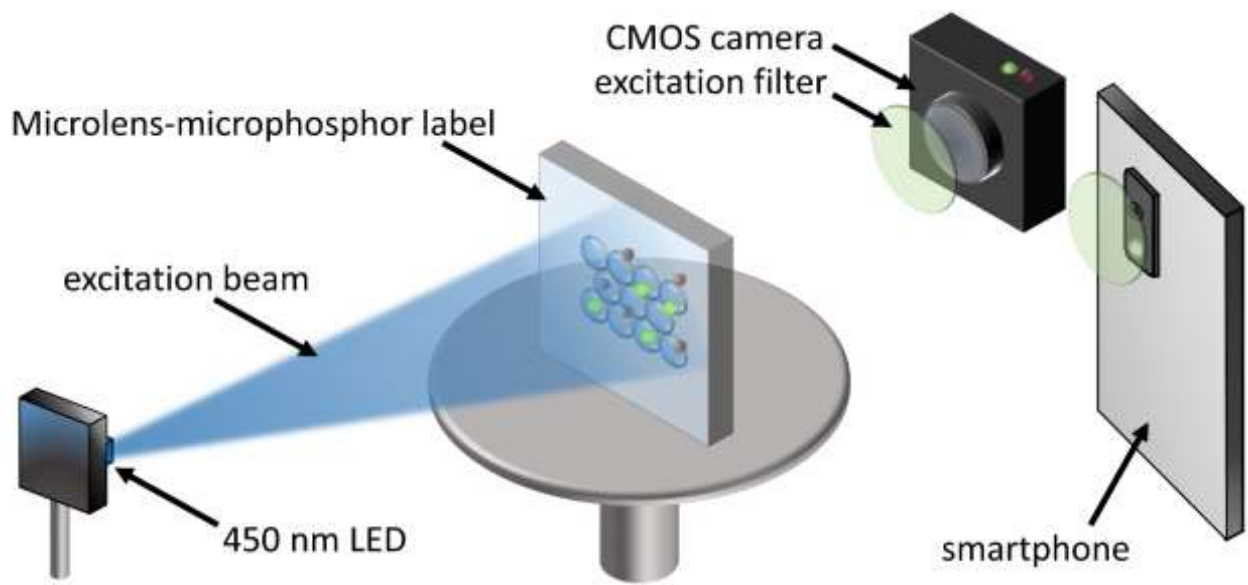

*Fig. S7. Schematic authentication setup including a 450 nm LED as an excitation source to the microlens-microphosphor label, and excitation filtered smartphone as well as a scientific camera as detectors*